The fraction of light nuclei in the mass composition of primary cosmic radiation at $E_0$=1-100 PeV

S.E. Pyatovsky[1], V.S. Puchkov

P.N. Lebedev Physical Institute, The Russian Academy of Sciences, Russia

The characteristics of extensive air showers (EAS) that are sensitive to the mass composition of primary cosmic radiation (PCR) are recorded in the EAS core. These characteristics include the parameters of the phenomenon of "halo" (diffuse darkening spots on the X-ray emulsion film (XREF)). Studies have shown that the bulk of the events recorded in XRECs (X-ray emulsion chambers) are produced by protons and He nuclei, which allows the estimation of the fraction of light nuclei in PCR.

The events under analysis were obtained in the XREC PAMIR experiment, and the properties of these events were studied at distances extending up to about 10 cm from the EAS axis with a resolution of approximately 30 $\mu$m. The investigation of the parameters of events in EAS cores by the halo method permitted analyzing the mass composition of PCR at $E_0$=10 PeV and assessing the fraction of light nuclei in PCR, which depend only slightly on the model of EAS propagation through the atmosphere. From an analysis of $\gamma$-ray families featuring a halo and multicenter halos obtained in the PAMIR XREC, it was found that the fraction of the $p$+He component of PCR was about 40%. The properties of proton induced EAS events were studied along with the KASCADE-Grande experimental data. On the basis of studying the dependence of the EAS age and $N_\mu$ on $N_e$ it is concluded that PCR mass composition becomes heavier in the range of $E_0$=1-100 PeV.

INTRODUCTION

The mass composition of PCR still remains the subject of discussions. The proton fraction in the PCR mass composition at primary energies in the range of $E_0$=1-100 PeV and the $p$+He fraction are estimated, respectively at 5 to 20% and at up to 70%, depending on the underlying experiment and on the model used to describe the propagation of EAS through the atmosphere. In order to assess the PCR mass composition, one determines and analyzes EAS parameters exhibiting minimum fluctuations in the course of the development of nuclear-electromagnetic cascades in the atmosphere. These parameters characterize events in EAS cores.

The detection of events in the vicinity of the EAS axis at distances of about several tens of centimeters is accomplished with the aid of resistive plate counters (RPC) in the Argo-YBJ experiment and XREC in the PAMIR experiments, and the experiments of the Brazilian-Japan Collaboration. The application of RPC with the aim of localizing the EAS axis permitted reaching a model accuracy of about 10 cm, but it turned out to be impossible to study the structure of the EAS core at these distances. The application of XREC is the only method that makes it possible to study the structure of events in the EAS core with a resolution of about 30 $\mu$m. Events that are detected by XREC are mostly, more than 96%, generated by the $p$+He component. The properties of $\gamma$-ray families, in particular, halo, are maximally sensitive to the PCR mass composition in view of the localization of $\gamma$-rays near the EAS axis.

Previously, it was assumed the PCR mass composition at an energy of $E_0$~3 PeV was dominated by the $p$+He component. Presently, however, the results of a number of experiments suggest that the PCR mass composition becomes heavier starting from $E_0$~1 PeV. Fig. 1 shows the $E_0$ spectrum of PCR in the knee region according to the EAS-TOP, Tibet III, IceTop, Tunka, Akeno, KASCADE-Grande, BLANCA, GAMMA, HiRes II, ARGO-YBJ, DICE, and CASA-MIA data. The results of the EAS-TOP and MACRO hybrid experiments, the Tibet AS$\gamma$ and BASJE experiments, and the experiments of the CASA-MIA Collaboration revealed a decrease in the $p$+He fraction in the PCR mass composition in the region of the knee of the PCR energy spectrum at $E_0$~3 PeV. According to data from the KASCADE-Grande experiments, the fraction of protons in the energy range of $E_0$=1-100 PeV does not exceed 10%. The results of the ARGO-YBJ experiment showed that the $p$+He fraction begins to decrease at $E_0$~1 PeV, with the result that the PCR mass composition becomes heavier. From the data in Fig. 1, it follows that the estimated fractions of light nuclei and the change in it in the PCR mass composition with $E_0$ differ substantially. The greatest discrepancy between the experimental data in estimating the $p$+He fraction is observed at $E_0$~10 PeV. The $p$+He fraction is about 10% according to the ARGO-YBJ data and 60% according to the IceCube data. A 50% discrepancy is indicative of the need for a more reliable estimation of the $p$+He fraction in the PCR mass composition at $E_0$=10 PeV by a method that would be as weakly model-dependent as possible.

---

[1] vgsep@ya.ru

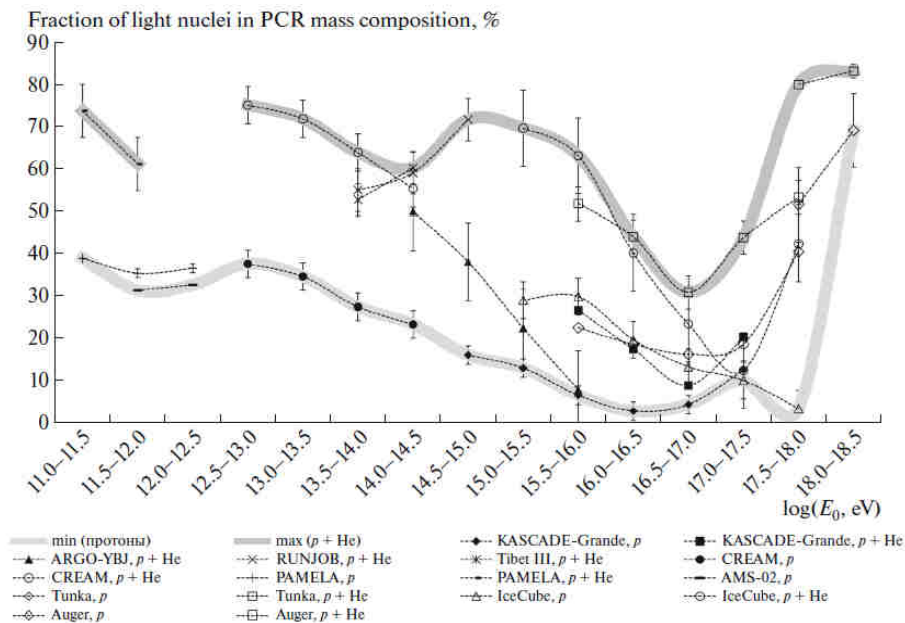

Fig. 1. Proton and $p$+He fractions in the mass composition of PCR according to data from the basic experiments.

Two basic aspects concerning the estimation of the fraction of light nuclei in the PCR mass composition were discussed in the studies of the KASCADE-Grande, Tunka, and Tunka-Rex Collaborations, as well as in some other studies, where LHCf data were also employed. The first consists in estimating the fraction of protons and helium nuclei in the PCR mass composition, while the second is the question of whether the PCR mass composition becomes heavier or lighter in the region of the knee of the PCR energy spectrum.

The halo method is applied in XREC experiments enabling the estimation of the $p$+He fraction on the basis of an analysis of events in the vicinity of the EAS axis. According to this method, the $p$+He fraction remains sizable at $E_0$=10 PeV. Fig. 2 shows the change in the proton and $p$+He fractions in the PCR mass composition over the energy range of $E_0$=1-100 PeV according to data of the XREC PAMIR, KASCADE, ARGO-YBJ, Tunka, and IceCube experiments. From the data in Fig. 2, it follows that the results of the XREC PAMIR experiment that concern the estimation of the proton and helium-nucleus fractions in the PCR mass composition at $E_0$=10 PeV on the basis of an analysis of halos are in good agreement with the results of the Tunka and IceCube experiments. The discrepancies between the proton and $p$+He fractions according to the XREC PAMIR, Tunka, and IceCube data and those according to the KASCADE and ARGO-YBJ data is 15% to 20%, which is due to the difference in the methods used in mass composition estimations to obtain and interpret features of events: one considers features of events within several centimeters from the EAS axis in the halo method but deals with features of events characterized by higher fluctuations far off the EAS core.

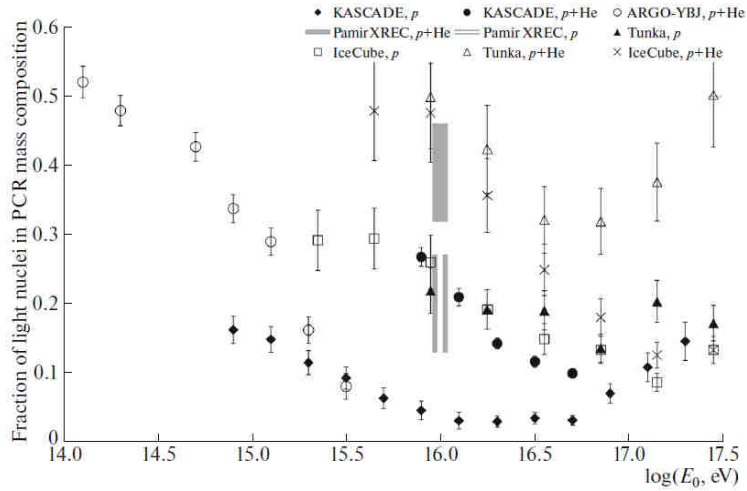

Fig. 2. Proton and p+He fractions in the PCR mass composition according to data from the XREC experiment PAMIR in comparison with the data of other experiments.

Basic experiments performed with XREC since the 1970s include the PAMIR (Tajikistan, 4300 m above sea level) and Tien-Shan high-mountain Scientific Station (Kazakhstan, 3400 m above sea level) experiments, the experiments of the Brazil–Japan Collaboration (Kambala, Tibet, 5400 m above sea level; Chacaltaya, Bolivia, 5280 m above sea level; and Fuji, Japan, 3700 m above sea level), and the JACEE and RUNJOB balloon-borne experiments. Only the XREC PAMIR experiment has so far given a statistically significant number of events (1294 $\gamma$-ray families and 61 halos).

Events detected in experiments with XREC are electromagnetic cascades induced by EAS and observed on an XREF as darkening points. These darkening spots were partitioned into four groups: (i) individual $\gamma$-rays associated with one to three closely lying darkening spots smaller than 1 mm$^2$ in area on XREF, (ii) $\gamma$-ray families (Fig. 3a) associated with numerous (several tens of) local darkening spots smaller than 1 mm$^2$ in area on XREF, (iii) single-center halos (Fig. 3b) or $\gamma$-ray families with halo as a diffusion darkening region of area corresponding to the halo criterion (in addition to the diffusion darkening region, individual $\gamma$-rays ($\gamma$-ray families) are observed in the halo image), and (iv) multicenter halos (Fig. 3c) as diffusion darkening regions on XREF that have areas whose sum satisfies the multicenter-halo criterion (in addition to these diffusion darkening regions, individual $\gamma$-rays ($\gamma$-ray families) are observed in the halo image).

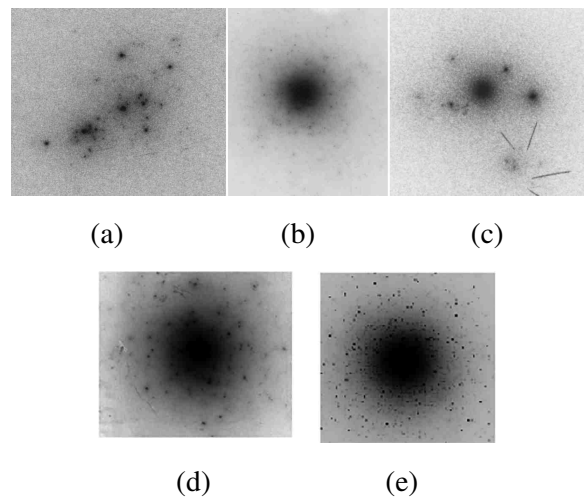

(a)  (b)  (c)

(d)  (e)

Fig. 3. (a) $\gamma$-ray family, (b) Tajikistan single-center halo, (c) multicenter halo, and $\gamma$-families with halo with the areas of ~1000 mm$^2$ and $E_0$~100 PeV: (d) experimental halo and (e) calculated halo.

For primary energies in the region of $E_0 \geq 0.1$ PeV to which shower arrays were rated, the estimates of $E_0$ and of the PCR mass composition depend on special features of EAS reconstruction in the atmosphere. At the same time, experiments with XREC arranged deep in the atmosphere detect proton-induced events. The greater the depth of the atmosphere, the higher the value of $E_0$ and the lighter the PCR nuclei that formed EAS and which XREC record as $\gamma$-ray families. In what concerns the estimation of the PCR mass composition, XREC provide the only weakly model-dependent experimental method for $E_0 \geq 0.1$ PeV that makes it possible to analyze events in EAS initiated by protons and helium nuclei. A limit on $E_0$ in experiments with XREC is determined by the selection criterion for $\gamma$-ray families with $\Sigma E_\gamma \geq 0.1$ PeV.

ESTIMATION OF THE $p$+He FRACTION IN THE PCR MASS COMPOSITION

We denote the PCR intensity at the EAS generation point by $I_0 = I_p + I_{He} + I_{>He}$, where $I_p$, $I_{He}$, and $I_{>He}$ are the intensities of primary protons, helium nuclei, and nuclei heavier than helium, respectively. Let $\tilde{N}_p$, $\tilde{N}_{He}$, and $\tilde{N}_{>He}$ be the numbers of $\gamma$-ray families observed on XREF if all of the PCR nuclei are protons, helium nuclei, and nuclei heavier than helium, respectively, and let $W_p$, $W_{He}$, and $W_{>He}$ be the probabilities for the formation of $\gamma$-ray families by protons, helium nuclei, and nuclei heavier than helium, respectively. We then have $I_0 = \tilde{N}_p / W_p = \tilde{N}_{He} / W_{He} = \tilde{N}_{>He} / W_{>He}$.

The number of the $\gamma$-ray families obtained experimentally is $N_0 = N_p + N_{He} + N_{>He}$, where $N_p$, $N_{He}$, and $N_{>He}$ are the numbers of $\gamma$-ray families produced by protons, helium nuclei, and nuclei heavier than helium, respectively.

We denote by $P_p$, $P_{He}$, and $P_{>He}$ the fractions of protons, helium nuclei, and nuclei heavier than helium, respectively. We then have $N_p = I_0 P_p W_p$, $N_{He} = I_0 P_{He} W_{He}$, and $N_{>He} = I_0 P_{>He} W_{>He}$.

Taking into account the relation $P_p + P_{He} + P_{>He} = 1$, we obtain

$$P_p = \left\{ \frac{N_0 W_p}{\tilde{N}_p} - W_{He}(1 - P_{>He}) - W_{>He} P_{>He} \right\} (W_p - W_{He})^{-1} \tag{1}$$

$$P_{He} = \left\{ \frac{N_0 W_p}{\tilde{N}_p} - W_p(1 - P_{>He}) - W_{>He} P_{>He} \right\} (W_{He} - W_p)^{-1} \tag{2}$$

From Eqs. (1) and (2), it follows that $P_p$ and $P_{He}$ depend on $W_p$, $W_{He}$, $W_{>He}$, and $N_0 / \tilde{N}_p$. The probabilities for the formation of halos by PCR nuclei are given in Table 1. From the data in Table 1, it follows that the probabilities $W_p$, $W_{He}$, and $W_{>He}$ differ by a factor of four or more, and this makes it possible to assess the fractions $P_p$, $P_{He}$, and $P_{>He}$. In Table 1, $W^{(100)}_{p,He,>He}$, $W^{(400)}_{p,He,>He}$, and $W^f_{p,He,>He}$ are the probabilities for the formation of, respectively, 100-TeV $\gamma$-ray families, 400-TeV $\gamma$-ray families, and structured halos with respect to $I_{p,He,>He}$. In the energy range of $E_0 = 5$-10 PeV, we have $W^{(100)}_p \cong 5.05\%$, $W^{(100)}_{He} \cong 0.79\%$, and $W^{(100)}_{>He} \cong 0.05\%$ (more than 96% of all $\gamma$-ray families for which $\Sigma E_\gamma \geq 0.1$ PeV are formed by protons and helium nuclei).

Table 1. Probabilities for the formation of $\gamma$-ray families by PCR nuclei (the EAS reconstruction criteria correspond to XREC PAMIR, $E_0 \geq 5$ PeV (this the halo-formation threshold)).

| $W_p$, % | | | | $W_{He}$, % | | | | $W_{>He}$, % | | | |
|---|---|---|---|---|---|---|---|---|---|---|---|
| $W_p^{(100)}$ | $W_p^{(400)}$ | Halo | $W_p^f$ | $W_{He}^{(100)}$ | $W_{He}^{(400)}$ | Halo | $W_{He}^f$ | $W_{>He}^{(100)}$ | $W_{>He}^{(400)}$ | Halo | $W_{>He}^f$ |
| 9.24 | 2.32 | 1.76 | 0.73 | 3.28 | 0.71 | 0.44 | 0.18 | 2.24 | 0.28 | 0.13 | 0.07 |

Fig. 4 shows the probabilities for the formation of halos by protons, helium nuclei, and nuclei heavier than helium versus $E_0$. From Fig. 4, it follows that up to $\lg E_0 = 16.7$, almost all of the halos are produced by protons and helium nuclei. The formation of halos by nuclei heavier that helium begins from $E_0 > 100$ PeV.

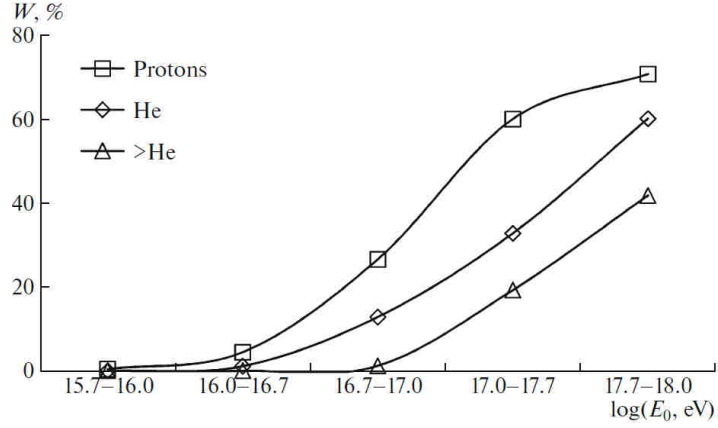

Fig. 4. Probability for the formation of halos by protons, helium nuclei, and nuclei heavier than helium versus $E_0$.

In order to assess the dependence of the halo method on the model variables $W_p$, $W_{He}$, $W_{>He}$, and $\tilde{N}_p$ in estimating $P_p$ and $P_{He}$, we assume that $W_p=nW_{He}$, $W_{He}=mW_{>He}$, and $w=N_0/\tilde{N}_p$:

$$P_p = \frac{wn-1}{n-1} + \frac{(m-1)}{(n-1)}\frac{P_{>He}}{m} \qquad (3)$$

$$P_{He} = \frac{n}{1-n}(w-1) - \frac{(1-mn)}{(1-n)}\frac{P_{>He}}{m} \qquad (4)$$

At $m=n=1$, the EAS reconstruction model is not sensitive to the type of the PCR primary particle; therefore, we do not apply it in describing the data from the XREC PAMIR experiment. As $m$ and $n$ grow, the features of halos (for example, halo statistics) produced by different nuclei come to be markedly different. Our calculations revealed that, upon the replacement of all PCR nuclei by protons, helium nuclei, or nuclei heavier than helium, the number of halos, $N_0$ would be, respectively, 140, 35, and 5 to 10. The boundary values of $N_0\pm\sqrt{N_0}=53-69$, $\tilde{N}_p\pm\sqrt{\tilde{N}_p}=128-152$, and $n=m=2$ determine the limiting possible values of $P_p$, $P_{He}$, and $P_{>He}$. The condition $n=m\geq 2$ (substantial difference in the halo-formation probability) indicates that the EAS reconstruction model is highly sensitive to the primary-nucleus type. In particular, the calculated ratio $W^{(100)}_{p,He,>He}$ (Table 1, XREC PAMIR) lies at the minimum possible boundary of the applicability of the halo method.

A feature peculiar to the XREC PAMIR experiment is that it detects predominantly $\gamma$-ray families generated by protons and, to a less extent, by helium nuclei (more 96% in all). For $m$, $n\gg 1$, it follows from Eqs. (3) and (4) that $P_{p,He}(n)/P_{p,He}(n+1)\to 1$ and $P_{p,He}(m)/P_{p,He}(m+1)\to 1$. At $n=m\geq 3.5$, for example, $P_{p,He}(n)/P_{p,He}(n+1)$ and $P_{p,He}(m)/P_{p,He}(m+1)$ both lie in the range of (0.9,1.1); at higher values of $n$ and $m$, the halo method becomes weakly model-dependent.

From Eqs. (1) and (2), it follows with allowance for the values in Table 1 for halos that $P_{>He}\leq 61\%$. For $P_{>He}>61\%$, no halo statistics will be present in XREC PAMIR. With allowance for $N_0\pm\sqrt{N_0}$, observed halo statistics (61 events within an exposure of $ST\cong 3000$ m² yr sr) requires at least a (39±6)% p+He fraction in PCR and at most a (61±6)% fraction of nuclei heavier than helium. Among these 39%, the minimum proton fraction is 20%, while the fraction of helium nuclei is not greater than 19% – a higher fraction of helium nuclei would not provide halo statistics observed experimentally. From Eqs. (1) and (2), it also follows that each halo detected experimentally increases the minimum p+He fraction by about 1%.

The halo method for assessing the PCR mass composition is characterized by (i) vast halo statistics – in contrast, for example, to statistics of multicenter halos (XREC PAMIR recorded 61 halos, including 14 mul-ticenter halos); (ii) a reliability of visual halo detection – in contrast, for example, to what we have for 400-TeV $\gamma$-ray families, whose statistics also depends on the method of measurement of $E_\gamma$; (iii) the detection of halos in XREC that are predominantly produced by protons and helium nuclei; and (iv) substantially different probabilities for the emergence of experimental events induced by protons, helium nuclei, and nuclei heavier than helium, which are used to estimate the PCR mass composition.

The XREC PAMIR array operates as a separator of EAS initiated by protons and helium nuclei. Knowing halo statistics and considering that the halos in XREC originate predominantly from protons and helium nuclei, we have estimated the minimum proton and helium-nucleus fractions in the PCR mass composition that provide visually observed statistics of halo events.

ESTIMATION OF THE $p$+He FRACTION ON THE BASIS OF MULTICENTER HALOS

Halo statistics; probabilities for halo formation by protons, helium nuclei, and nuclei heavier than helium (according to the data in Table 1, they are several fold different); and the fraction of multicenter halos are halo features that are sensitive to the PCR mass composition.

We denote by $N_S$ the number of experimental multicenter halos and by $S_0$ the experimental fraction of multicenter halos; we have $N_S=N_0 S_0$ ($N_0=61\pm\sqrt{61}$, so that $N_S=14\pm\sqrt{14}$). The fractions $S_0$ associated with protons, helium nuclei, and nuclei heavier than helium are given in Table 2 along the experimental value of $S_0$ in the XREC PAMIR experiment.

Table 2. Fraction of multicenter halos produced by protons and helium, carbon, and iron nuclei and respective experimental fractions according to XREC PAMIR data.

| $S_{0\,p}$ | $S_{0\,He}$ | $S_{0\,C}$ | $S_{0\,Fe}$ | $S_{0\,PAMIR}$ |
|---|---|---|---|---|
| 0.25 | 0.45 | 0.59 | 0.70 | 0.23±0.07 |

As applied to multicenter halos, Eqs. (1) and (2) yield

$$P_p = \left\{\frac{N_0 W_p}{\bar{N}_p} S_0 - W_{He} S_{He}(1 - P_{>He}) - W_{>He} P_{>He} S_{>He}\right\} (W_p S_p - W_{He} S_{He})^{-1} \quad (5)$$

$$P_{He} = \left\{\frac{N_0 W_p}{\bar{N}_p} S_0 - W_p S_p(1 - P_{>He}) - W_{>He} P_{>He} S_{>He}\right\} (W_{He} S_{He} - W_p S_p)^{-1} \quad (6)$$

With allowance for the values in Table 2 for multicenter halos – $S_0$=0.16-0.3 – the dependences in Eqs. (5) and (6) lead to the following estimates: $P_{>He}\leq 57\%$ and $p$+He$\geq 43\%$.

ESTIMATION OF THE CHANGE IN THE PCR MASS COMPOSITION

The number of muons, $N_\mu$, and the EAS age, $S$, were the features of EAS that were used to estimate the change in the PCR mass composition with $E_0$. It turns out that $N_\mu$ grows with increasing $N_e$ and $A$, while $S$ grows with increasing $A$ and becomes smaller with increasing $N_e$. The quantities $N_\mu$ and $S$ were analyzed along with KASCADE-Grande data (1 million events from KASCADE database).

In the range of lg$N_e$=6.0-6.5, $\Delta\langle S\rangle\cong 0$ – the reduction of $\langle S\rangle$ with increasing $N_e$ is compensated by the growth of $\langle S\rangle$ with increasing $A$. At lg$N_e$=6.0, $\langle A\rangle$ should not exceed 35 (Si group). In the range of lg$N_e$=4.5-5.0, $\Delta$lg$A\cong 0$ ($\langle A\rangle$ does not change), the values of $\langle A\rangle$ in Table 3. In the energy range of $E_0$=1-100 PeV, the PCR mass composition remained mixed, with the $\langle S\rangle$ value corresponding to nuclei of the CNO group for KASCADE-Grande data; the PCR mass composition becomes heavier with increasing $N_e$; in the knee region of the $E_0$ spectrum of PCR, $\langle S\rangle$ does not change – the growth of $S$ with increasing $A$ compensates for the reduction $S$ with increasing $N_e$.

Table 3. Change in <A> with $N_e$.

| lg$N_e$ | 4.5 - 5.0 | 5.0 - 5.5 | 5.5 - 6.0 | 6.0 - 6.5 | 6.5 - 7.0 | 7.0 - 7.5 | 4.5 - 7.5 |
|---|---|---|---|---|---|---|---|
| ΔlgA | -0.018 | 0.157 | 0.195 | 0.139 | 0.034 | 0.165 | 0.672 |

In the case of protons and iron nuclei, the dependence of $N_\mu$ on $E_0$ has the form

$$\lg N_\mu^p = (0.86\pm0.01)\lg(E_0[\text{ПэВ}])+(3.61\pm0.01), R_a^2=0.999 \qquad (7)$$

$$\lg N_\mu^{Fe} = (0.85\pm0.01)\lg(E_0[\text{ПэВ}])+(3.86\pm0.01), R_a^2=0.999$$

$N_\mu$ grows with increasing $A$ as $N_\mu \sim A^\alpha$. Taking into account Eqs. (7), we obtain $\alpha=0.14$, which corresponds to the effective multiplicity $N=19$ of $\pi^{0,\pm}$ production. At $E_0=10$ PeV, values in the range of <A>=8-9 correspond to the difference $\Delta\lg N_\mu=0.12-0.13$. The change in <A> with $E_0$ is given in Table 4. An analysis of the data on the distribution of $N_\mu/E_0$ shows that the PCR mass composition at $E_0=10$ PeV remains mixed, its average nuclei not being heavier than nuclei of the CNO group.

Table 4. Dependence of <A> on $E_0$.

| $E_0$, ПэВ | 2 | 4 | 9 | 18 | 35 |
|---|---|---|---|---|---|
| <A> | 10±2 | 17±4 | 9±2 | 5±1 | 7±1 |

Figure 5 shows the experimental dependences $N_\mu(N_e)$ (according to the EAS KASCADE-Grande database) and the dependences $N_\mu(N_e)$ obtained for protons from the experimental EAS KASCADE-Grande database. From Fig. 5, it follows that $N_\mu^{EAS}$ grows with $N_e$ faster than $N_\mu^p$ does, which is indicative of the growth of the mass of particles in PCR.

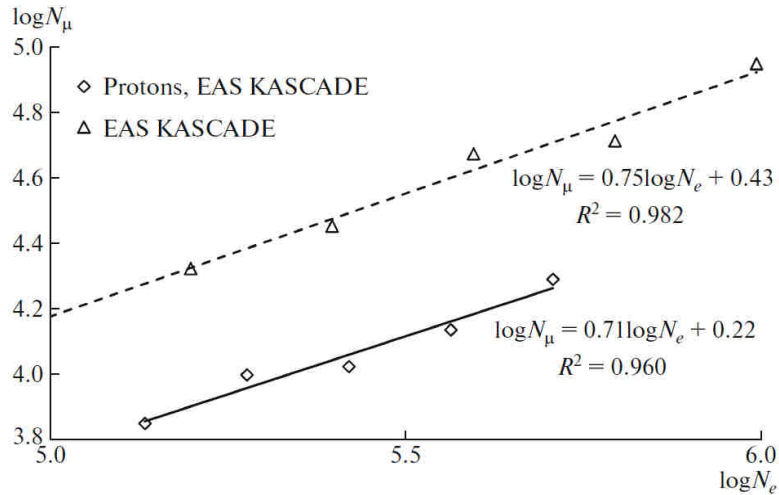

Fig. 5. Dependence of $N_\mu$ on $N_e$ in the EAS KASCADE-Grande data. The black solid curve and dashed curve represent, respectively, the results for protons and the EAS KASCADE data.

CONCLUSIONS

The application of the halo method to the XREC PAMIR database has revealed that, for ensuring observed statistics of $\gamma$-ray families with halo, the $p$+He fraction in the PCR mass composition at $E_0=10$ PeV should not be less than (39±6)%. According to data from the KASCADE-Grande experiments, the PCR mass composition becomes heavier in the energy range of $E_0=1-100$ PeV.

Our present analysis has led to the following conclusions:

(i) The estimate obtained here for the *p*+He fraction in the PCR mass composition is a lower bound. The *p*+He fraction will increase upon taking into account additional conditions – for example, the change in the relationship between $P_p$, $P_{He}$, and $P_{>He}$ with $E_0$.

(ii) Within the halo method, one employs events detected near the EAS axis, which carry information about the primary interaction of PCR nuclei with nuclei of the atmosphere.

(iii) The probabilities for halo formation by protons and helium nuclei does not differ several fold, which renders the halo method weakly model-dependent. Further, the XREC PAMIR array is viewed as a separator of protons and, to a less extent, as a separator of helium nuclei.

(iv) The halo-formation threshold is substantially higher than $E_0=0.1$ PeV. Owing to this, the halo method is applicable in the $E_0$ region where other methods for assessing the PCR mass composition are mostly indirect and model-dependent.